\documentclass{aa}
\usepackage{ffbib}
\usepackage{psfig}
\usepackage{epsfig}
\def\gsim{\;\lower4pt\hbox{${\buildrel\displaystyle >\over\sim}$}\,}
\def\lsim{\;\lower4pt\hbox{${\buildrel\displaystyle <\over\sim}$}\,}

\def \xmm {{\em XMM-Newton}}
\def \ch {{\it Chandra}}
\def \src {XMMU J061705.5+222130}

\def \ergsec{\hbox{erg s$^{-1}$}}
\def \ergcms{\hbox{erg cm$^{-2}$ s$^{-1}$}}
\def \hcm {\hbox {\ifmmode $ atom cm$^{-2}\else atom cm$^{-2}$\fi}}
\def \arcmin {\hbox{$^\prime$}}
\def \arcsec {\hbox{$^{\prime\prime}$}}

\def\approxgt{\mathrel{\hbox{\rlap{\lower.55ex \hbox {$\sim$}}
        \kern-.3em \raise.4ex \hbox{$>$}}}}
\def\approxlt{\mathrel{\hbox{\rlap{\lower.55ex \hbox {$\sim$}}
        \kern-.3em \raise.4ex \hbox{$<$}}}}

\begin{document}

%\thesaurusECK: 08.19.5; 09.09.1: \src; 09.19.2; 13.25.4)}  

\title{The plerion nebula in IC~443: the XMM-Newton view}

%\thanks{Based
%on observations obtained with \xmm, an ESA science mission with instruments
%and contributions directly funded by ESA Member States and the USA (NASA)}}
%\subtitle{I. A study of the plerion nebula}

      \author{F. Bocchino\inst{1,2}
         \and A. M. Bykov\inst{3}
}
\offprints{F. Bocchino (fbocchin@estec.esa.nl)}

\institute{
       Astrophysics Division, Space Science Dept. ESA-ESTEC,
              Postbus 299, 2200AG Noordwijk, The Netherlands
\and
       Osservatorio Astronomico di Palermo, Piazza del Parlamento 1,
       90134 Palermo, Italy
\and
       A.F. Ioffe Institute for Physics and Technology,  
           St. Petersburg, Russia, 194021
}

\date{Received 10 April 2001;Accepted 30 May 2001}

\abstract{\xmm ~observations of the X-ray feature 1SAX J0617.1+2221 in
the IC~443 supernova remnant are reported.  We resolve the structure of the
nebula into a compact core with a hard spectrum of photon index $\gamma =
1.63^{+0.11}_{-0.10}$ in the 2--10 keV energy range. The nebula also has
an extended ($\sim 8\arcmin \times 5\arcmin$) X-ray halo, much
larger than the radio emission extension.  The photon index softens,
following a linear scaling with distance from the centroid, similar to
other known X-ray plerions.  The index range is compatible with
synchrotron burn-off models. All the observational evidence points
toward a confirmation of the plerionic nature of the nebula, as
recently suggested by a \ch\  observation, but with characteristics
more similar to ``non Crab-like" plerions.  We discuss the implications
on the synchrotron nebula magnetic field if the $>100$~MeV emission
reported by {\it CGRO EGRET} is produced by the synchrotron emission.
We also constrain the thermal emission of the central object, arguing that
the surface temperature should be around 0.1 keV, although other
possible fits cannot be excluded on the base of the \xmm\  data.
 }

\maketitle

\markboth{Bocchino \& Bykov}
{The pulsar nebula in IC~443}

\section{Introduction}

The processes of conversion of pulsar spin-down energy into high-energy
emission in pulsar wind nebula are of great physical interest (e.g.,
\cite{kc84}; \cite{che00}).  Modern arcsec resolution instruments such
as on \xmm\ and \ch\  that are sensitive to photons up to 10 keV provide
unique possibilities to study pulsar nebulae, because the non-thermal
emission of the nebulae may be easily detected and studied in this band
(e.g.,  Vela-X, \cite{hgh01}, \cite{pzs01}; G21.5-0.9, \cite{scs00},
\cite{wbb01}; and 3C58, \cite{bwm01}), even when immersed in the soft X-ray
emission of the companion shell, as is often the case.

Indeed, the shell SNR IC~443, once though to be mostly thermal in the
X-ray band (\cite{pss88}; \cite{aa94}) has been discovered to emit hard
X-ray emission by \cite*{wah92}.  ASCA Gas Imaging Scintillator (GIS) 
observations by
\cite*{kpg97} discovered the localized character of the hard  X-ray
emission and its non-thermal nature. They concluded that most of the
2--10~keV photons came from an isolated emitting feature and
from the South East elongated ridge of hard emission. \cite*{pre99} and
\cite*{bb00} reported a hard component detected with the
Phoswich Detector System (PDS) on BeppoSAX
and two compact X-ray sources corresponding to the ASCA sources
detected with the BeppoSAX Medium-Energy Concentrator Spectrometer
(MECS) (1SAX J0617.1+2221 and 1SAX J0618.0+2227).
Very recently, 1SAX J0617.1+2221 has been observed by \ch\ as reported
by \cite*{ocw01}, who also show a VLA observation at 1.46, 4.86 and
8.46 GHz, and a polarization measurement.  They argue that the hard
radio spectral index, the amount of polarization and the overall X-ray and
radio morphology strongly suggest that the source is a plerion nebula
with a point source in it, whose characteristic cometary shape is due
to supersonic motion of the neutron star.  However, the limited
counting statistics of the 10 ks \ch\  observation do not allow a
detailed spectral study of the nebula, which is required to compare
this new plerion with current theoretical models. Moreover, IC~443 is a
possible candidate for the CGRO EGRET $\gamma$-ray source 3EG
J0617+2238 (\cite{hbb99}) having a flux above 100 MeV of $\sim 5 \times
10^{-7}$ ph s$^{-1}$ cm$^{-2}$ with a photon index of $2.01 \pm 0.06$.
More detailed measurements of the nebula spectral properties are needed
to study the relation between 3EG J0617+2238 and the nebula.

%Supernova remnant IC~443 (G189.1+3.0) is one of the best objects to
%study rich phenomena accompanying supernova interaction with a
%molecular cloud (e.g. \cite{bgb88}; \cite{vjp93}). IC~443 was a target
%of X-ray observations with { \it HEAO 1} (\cite{pss88}), {\it Ginga}
%(\cite{wah92}), {\it ROSAT} (\cite*{aa94}), {\it ASCA} (\cite{kpg97},
%hereafter K97) \sax\ (\cite{pre99}; \cite{bb00}) and
%\ch\ (\cite{ocw01}).

%The soft X-ray 0.2--3.1 keV surface brightness map of IC~443 from the
%{\it Einstein} Observatory (\cite{pss88}) shows bright features in the
%northeastern part of the remnant as well as bright soft emission from
%the source central part.  The presence of nearly uniform X-ray emission
%from the central part of the remnant was clearly seen also by {\it
%ROSAT} (Asaoka \& Aschenbach, 1994), and corresponds to the emission
%from hot (T$\sim 10^7$ K), low density gas interior to the shock.  

In this {\it paper} an \xmm\ study of the recently discovered plerion nebula 
in IC~433 is presented. In particular, we use the large
effective area of the EPIC instrument to address the synchrotron
burn-off effect in the nebula, to resolve its structure and measure the
flux, and to constrain the thermal radiation of the central object in
the nebula.

\section{Observations}

IC~443 was observed as part of the Cal/PV phase of the \xmm\
Observatory (\cite{jla01}). Here, we have used the two
observations centered on
$6^h17^m24^s.3$ $+22^d26^m43^s$ (J2000) performed on 2000 September 27.
Data from the two MOS (\cite{taa01}) cameras and the PN (\cite{sbd01})
camera are used.  The MOS and PN cameras are CCD arrays which collect
X-ray photons between 0.1 and 15 keV and have a field of view of
$30^\prime$ diameter. The pixel size is $1.1\arcsec$ and $4.1\arcsec$ for MOS
and PN, respectively, while the mirror Point Spread Function is 6\arcsec\
Full-width at half maximum. 
The data were acquired with the medium
filter and in full image mode, and therefore the temporal resolution is
2.5~s and 73~ms for the MOS and PN, respectively. The poorer spatial
resolution of the PN is compensated for by its greater sensitivity, on the
average 20--30\% more than the combined MOS cameras.

The Standard Analysis System (SAS) software used (version
5.0.1, xmmsas-20001215) takes cares of most of the required events
screening. However, we have further screened the data to eliminate some
residual hot pixels and occasional background enhancement during
intervals of intense incident flux of soft protons. In particular, we
extracted the background lightcurve at energies $>10$~keV and
identified time intervals of unusually high count rates (typically
more than 1~s$^{-1}$) and removed them from subsequent analysis.
We merged the the event files of the subpointings
before continuing with the analysis. The total screened and unscreened
exposure times are 24 and 32 ks, respectively.

\section{Results}

\subsection{X-ray morphology}

\begin{figure}
  \centerline{\psfig{file=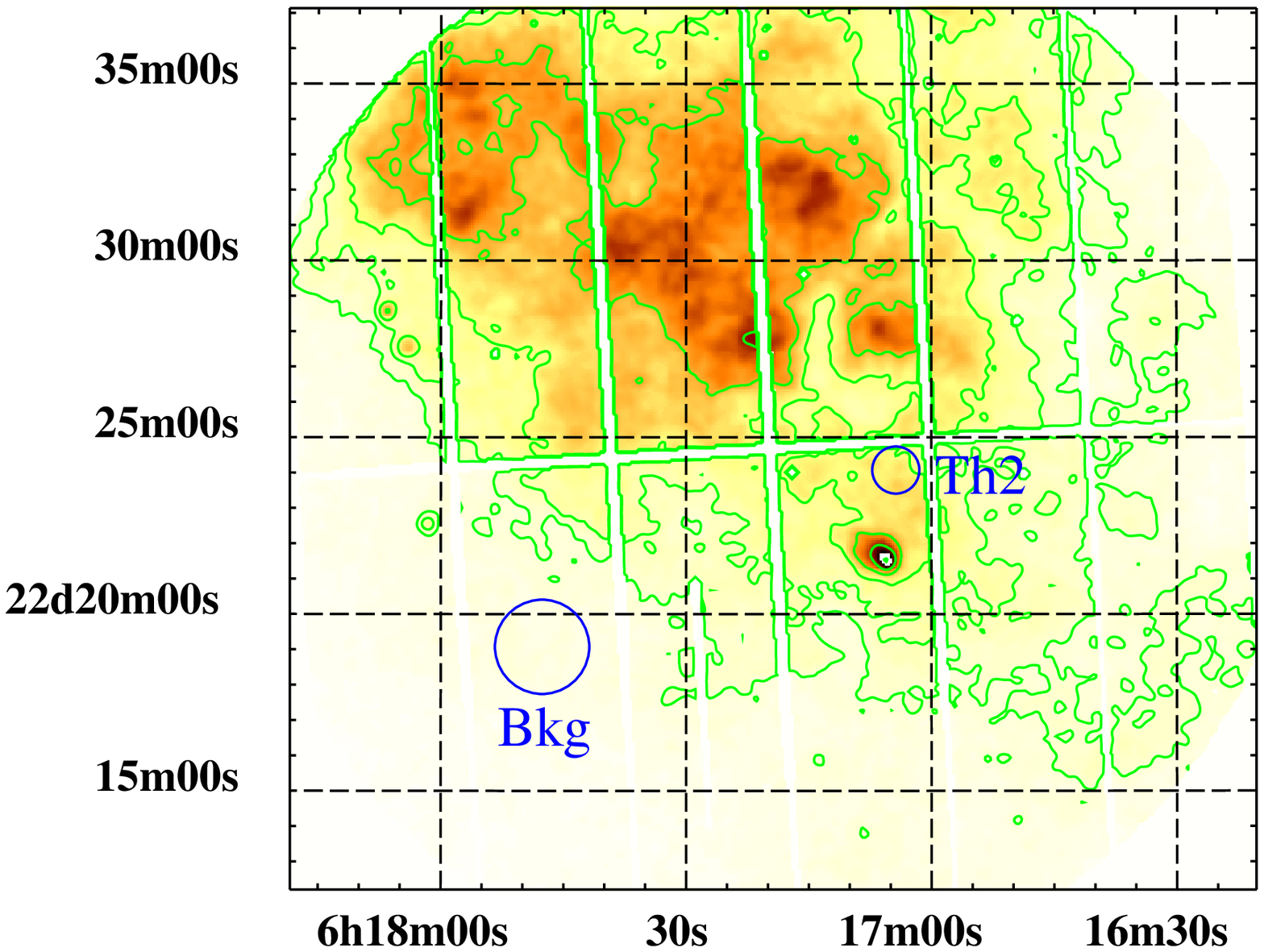,width=9.0cm}}
  \centerline{\psfig{file=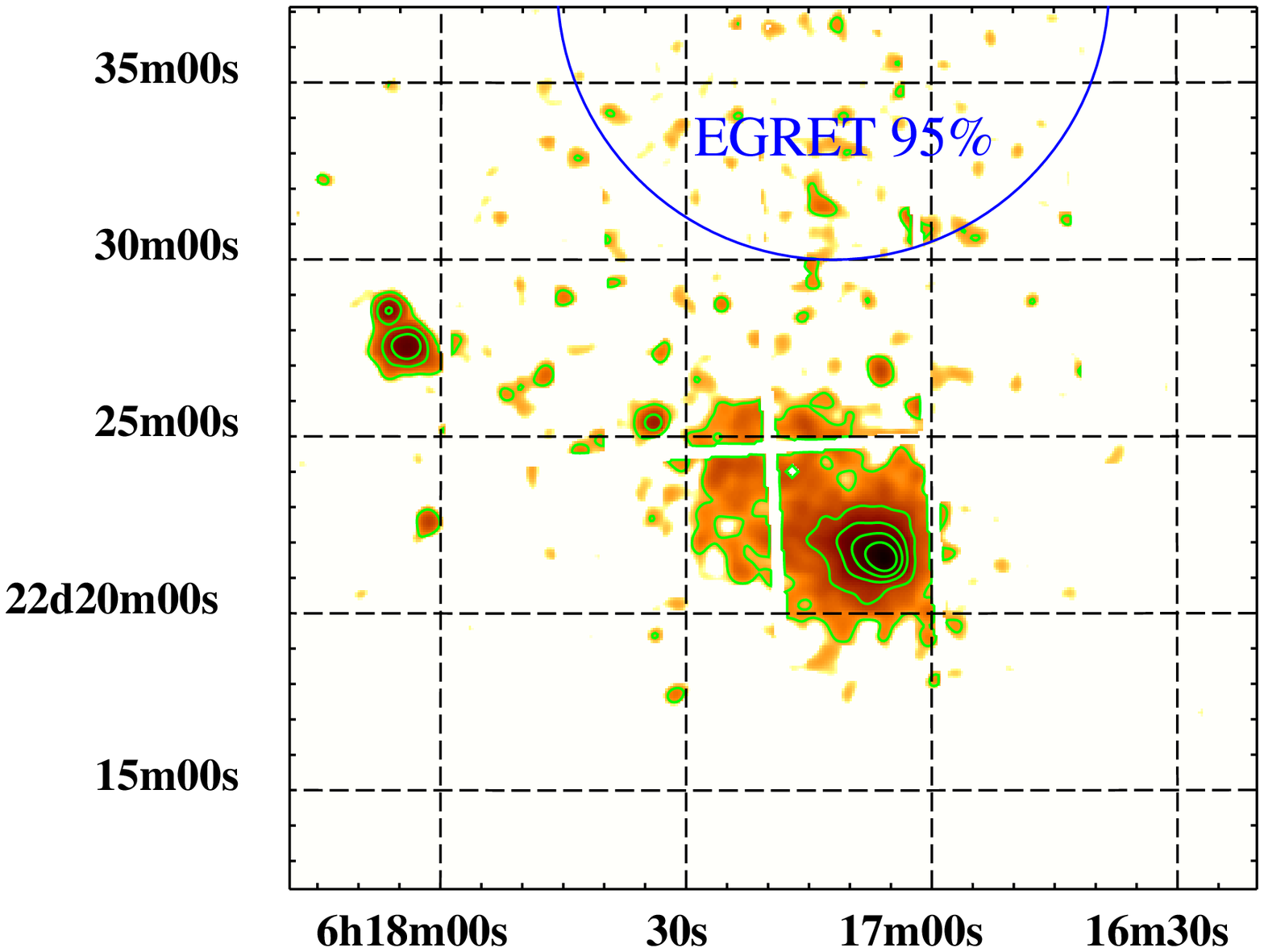,width=9.0cm}}
  \caption{4\arcsec\  pixel PN images of the IC~443 plerion. {\em Top:}
  soft energy band (0.5--2 keV), linear colorscale, logarithmic
  contours between 1/16 and the peak value, 30~count~pixel$^{-1}$, 
each contour is a
  factor of two from the previous and smoothed with a 1.5 pixel $\sigma$
  Gaussian. {\em Bottom:} Hard energy band (3--10 keV), log colorscale,
  same log contours spacing but between 1/32 and the peak at 12.5
  count~pixel$^{-1}$, smoothed with a 3 pixel $\sigma$ Gaussian. The 95\%
  confidence {\it EGRET} error circle for 3EG J0617+2238 is also shown}

  \label{pnimage}
\end{figure}

We extracted PN images with a 4\arcsec\ pixel size and exposure images
at the same resolution.  We defined two energy bands in which to
extract images, namely 0.5--2 keV (hereafter the soft band) and 3--10 keV
(hard band). Fig.~\ref{pnimage}, top panel, shows the 0.5--2~keV PN
image of the South Eastern section of IC~443. Most of the emission in this
band is of thermal origin, as established by \cite*{aa94} based on
ROSAT data.

Fig.~\ref{pnimage}, bottom panel, shows the 3--10 keV PN image of the
same field with a logarithmic color scaling and a heavier smoothing
chosen in order to emphasize the faint diffuse emission.  Most of the
thermal emission associated with IC~443 is not present in this band, and
therefore it is well suited to the study of any hard compact
X-ray sources in the vicinity of this SNR. Indeed, this image shows
several point sources, besides the plerion nebula itself.  We see that
the nebula is much more extended then reported by \cite*{ocw01} and can
be represented by an ellipse of $8\arcmin \times 5\arcmin$.

\begin{figure}
  \centerline{\psfig{file=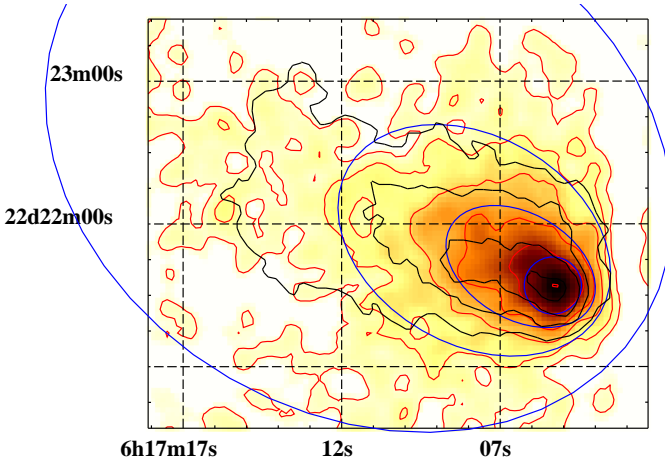,width=9.0cm}}
  \caption{MOS1+2 3--10 keV image closeup of the core of the nebula,
  smoothed with a Gaussian of 4\arcsec\  $\sigma$. Black contours are
  drawn from \protect\cite*{ocw01}, at 8.46 GHz, in steps of 1 mJy
  beam$^{-1}$.  The elliptical regions used in spectral analysis are
  also displayed, as well as logarithmic X-ray contours between 1/32
  and the peak value (8 count pixel$^{-1}$), each one a factor of two from
  the previous}

  \label{plerion}
\end{figure}

By fitting a two dimensional Gaussian to the image, we find that the
location of the plerion nebula centroid is $6^h17^m05.5^s$ $22^\circ
21\arcmin 30\arcsec$ (J2000, therefore \src), with an estimated
accuracy of 5\arcsec, due to both statistical uncertainties (at
$3\sigma$) and systematic errors introduced by attitude reconstruction,
and therefore in agreement with the position of the \ch\ source CXOU
J061705.3+222127.  Fig.~\ref{plerion} shows the combined MOS 3--10~keV image
of the nebula core.  The morphology of the faint rim matches somewhat
that of the bright core, which may suggest a common origin, in
agreement with the plerionic interpretation. Fig.~\ref{plerion} also
shows that the radio halo at 8.46 GHz observed by \cite*{ocw01} has
somewhat different shape, and is smaller than the X-ray full extent of
the nebula.

We also note that the source 1SAX~J0618.0+2227 is resolved with {\it
XMM-Newton} into two closeby sources (see Fig.~1). Moreover, there are
other hard X-ray sources in the field of view. Some of these sources
might be related to the SNR. The location of these sources is
consistent with a scenario concerning the interaction of the SNR with
molecular cloud, as proposed by \cite*{bb00}.

\subsection{Spectral analysis}

\subsubsection{Contamination from soft thermal emission}

The spectral analysis of the plerion is rather complicated because of
the presence of diffuse thermal emission on large spatial scales which is
unassociated with the plerion nebula. Fig.~\ref{pnimage} shows that
the plerion is partially superimposed on a large region of soft
emission, running from the North East to the 
South West, which links the plerion region with
the very bright soft region around $6^h17^m30^s$ $22^\circ 30\arcmin$.

In order to cope with these problems, we adopted the following
procedure. First, we identified a ``background area" (Bkg in Fig.
\ref{pnimage}) which is outside the thermal emission of IC~443. This
background have been corrected for vignetting and area effects prior to
subtraction of the source spectra. The background correctly includes
emission from G189.6+3.3, the ``companion" SNR identified by
\cite*{aa94}, which encompasses all the field of view of the \xmm\
observation. Then, we defined a sample of the thermal region in which
the plerion is partially immersed (Th2 in Fig. ~\ref{pnimage}).
Unfortunately, the Th2 region cannot be located outside the plerion
nebula, because the spectral properties may be different in a region
so far from the nebula.  The thermal region was fitted with the
two-temperature optically thin plasma model of \cite*{mgo85} and
\cite*{log95} with interstellar absorption of \cite*{mm83}, using the
appropriate PN response matrix for off-axis sources. An additional
power-law component was used to model the non-thermal emission of the
plerion falling inside the Th2 extraction region. The above 2T model
has been successfully used by \cite*{aa94} to describe the ROSAT
data of this part of IC~443. A proper study of the remnant thermal
emission including more appropriate models (e.g., Non-Equilibrium of
Ionization) is beyond the scope of our present work.

The results obtained from the thermal regions show that the two
temperature thermal model used by \cite*{aa94} provides a good
description of the data with best-fit parameters in agreement with
the findings of ROSAT ($kT_1=0.18^{+0.10}_{-0.03}$,
$kT_2=0.90^{+0.10}_{-0.05}$, $N_{\rm H}=8.0\pm1.5 \times 10^{21}$,
$\chi^2/dof=71/66$).

%A slightly better model is provided by a single temperature thermal model
%with variable abundances, with $kT=0.4$, abundances lower than solar,
%and a lower $N_{\rm H}$ value ($4-7\times 10^{21}$), but since this model
%does not fit the brighter region ``Pl3", we prefer the 2T model.  We
%have verified that a plane parallel Non-Equilibrium Ionization shock
%model (model {\sc pshock} in XSPEC) with standard abundances,
%do not give acceptable results. A fully analysis of the IC~443 thermal
%emission is beyond the scope of this work.

\subsubsection{The central compact source}

\begin{figure}
  \centerline{\psfig{file=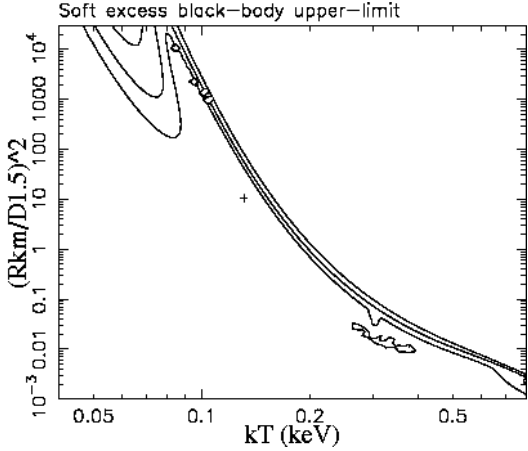,width=7.0cm}}
  \caption{Upper limit to the blackbody component in the spectral fits
  of \src. We present 68\%, 90\% and 99\% confidence level $\chi^2$
  contours for 4 interesting parameters versus the blackbody
  temperature and normalization (in units of $(R_{\rm km}/D_{1.5})^2$,
  where $R_{\rm km}$ is the blackbody radius in km and $D_{1.5}$ is
  the distance in units of 1.5 kpc. The best-fit values are marked with
  a cross}

  \label{bb}
\end{figure}

We analyzed the spectrum of the compact source \src, extracting its
spectrum from a 6\arcsec\ radius region centered on the source
centroid. In order to take into account the residual thermal emission
in the fittings, we used a power-law model together with the 2T
model used in the Th2 region to fit the data. The absorption,
temperatures and normalization ratio of the two thermal component were
kept fixed to the values found in Th2.  The best-fit results are
reported in Table~\ref{srcpar}.  Since the best-fit normalization of
the thermal component is a factor of 2 higher than that expected by
scaling the value found in Th2, we have tested for the presence of an
additional black-body component in the spectrum of this central part of
the nebula, by adding this component to the power-law instead of the 2T
component. There is no a significant improvement in $\chi^2$, but a
best-fit is found with kT=0.13 keV and a neutron star (NS) radius of 3.3
km. The corresponding values of the black-body temperature and radius
as observed at infinite distance allowed by our data are shown in 
Fig.~\ref{bb}.  At the temperature derived by \cite*{ocw01} with \ch, kT=0.7
keV, the upper-limit to the 1--5 keV unabsorbed flux of the BB
component is $4.4\times 10^{-13}$ erg cm$^{-2}$ s$^{-1}$, (consistent
with the \ch\ detection of a flux of $2\times 10^{-13}$ \ergcms), but
the high temperature and very low emitting area (even for heated polar
cap models) points against this region of the parameter space.

We have also investigated the presence of periodic signal by performing
an FFT of the PN events collected in the NS source region, i.e. a 
circular region
centered on the centroid with a 6\arcsec\  radius. We did not find a
periodic signal at the 99\% confidence level in the $10^{-4}-6.5$ Hz
range, with an upper limit, at the same confidence level, of 16.8\% of
sinusoidal pulsated fraction.

\subsubsection{The diffuse nebula}

In order to test the plerionic interpretation, we have tried to detect
the softening of the spectrum toward the outer regions of the nebula, a
well known effect which has been observed in other plerions (e.g. 3C~58,
\cite{tsk00} and \cite{bwm01}; G21.5-0.9 by \cite{scs00} and \cite{wbb01}) 
and which is caused by the short
lifetime of high energy electrons compared with those with lower
energies.  Given the peculiar shape of this nebula, we
extracted spectra from four elliptical annuli, shown in 
Fig.~\ref{plerion}, and fit them separately.  We chose the
ellipse sizes in order to minimize the point spread function 
(PSF) contamination effects. For
instance, the first region is a 12\arcsec\ radius circle, in which
$\sim$75\% of the PSF integral is included.

Because of the contamination by thermal emission, we have followed the
same approach used in fitting the NS region, i.e. a fixed parameter 2T
model and a power-law.  Whilst this procedure does not take into account
the uncertainties associated in the Th2 fit (which however are not
large), it provides an accurate way of not rejecting useful data,
thus reducing the overall uncertainties in the spectral parameter of
the nebula.  For comparison, we also show the results obtained with
fits to the hard band spectrum ($E>3$ keV) and a single power-law.

\begin{table}
\caption{Summary of PN spectral fitting results.}
\label{srcpar}
\medskip
\centering\begin{minipage}{8.7cm}
\begin{tabular}{lccccc} \hline
Name & $\gamma$\footnote{Fit to power-law only, 3--10 keV, $N_{\rm H}=8\times 10^{21}$ cm$^{-2}$.} &
$\gamma$\footnote{Fit to power-law + fixed parameters 2T components
($N_{\rm H}=8\times 10^{21}$, kT$_1=0.2$, kT$_2=0.9$, $norm_1/norm_2=10$.
Full bandwidth (0.5--10 keV).} & flux\footnote{Power-law unabsorbed flux in the
2--10 keV band.}$\times 10^{-13}$ & $\chi^2/dof$ \\

     & &  & erg cm$^{-2}$ s$^{-1}$ \\
\hline

NS     &            -           & $1.63^{+0.11}_{-0.17}$ & $6.8\pm1.7$ & 27/24 \\
PlCore & $1.71^{+0.44}_{-0.23}$& $1.63^{+0.11}_{-0.10}$ & $6.3\pm0.8$ & 40/48 \\
Pl1   & $1.95^{+0.35}_{-0.18}$& $1.81^{+0.08}_{-0.06}$ & $12.8\pm1.1$ & 119/108 \\
Pl2   & $1.88^{+0.31}_{-0.16}$& $1.88^{+0.07}_{-0.06}$ & $13.8\pm1.2$ & 149/149 \\
Pl3   & $2.70^{+0.31}_{-0.26}$& $2.30^{+0.09}_{-0.09}$ & $17.2\pm1.7$ & 451/435 \\

\noalign{\smallskip}
\hline
\end{tabular}
\end{minipage}
\end{table}

\begin{figure}
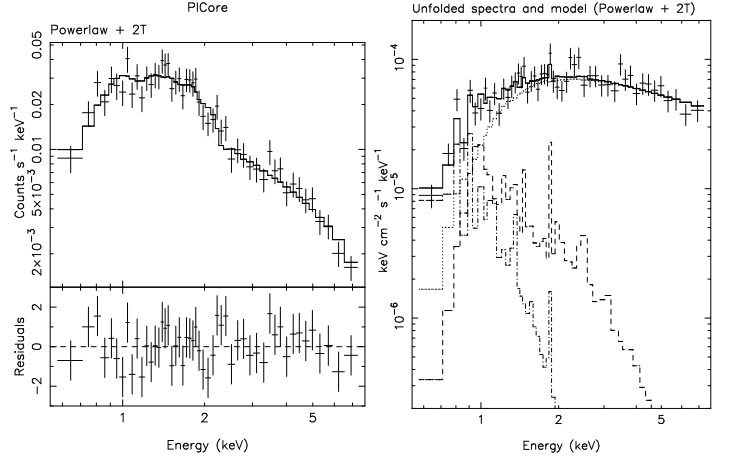

  \centerline{\hbox{
     \psfig{file=FIGURES/fig4a.ps,width=6.0cm,angle=-90}
     \psfig{file=FIGURES/fig4b.ps,width=6.0cm,angle=-90}
  }}
  \caption{Folded and unfolded PN spectrum of the central region of the
  plerion (within 12\arcsec\  from \src). The best-fit model (a
  power-law and two component thermal plasma with parameter derived from
  the Th2 region) is also shown. In the unfolded panel, the
  contribution of the thermal components (dashed and dot-dashed) and
  the power-law (dotted) are shown}

\label{plsp}
\end{figure}

\begin{figure}
  \centerline{\psfig{file=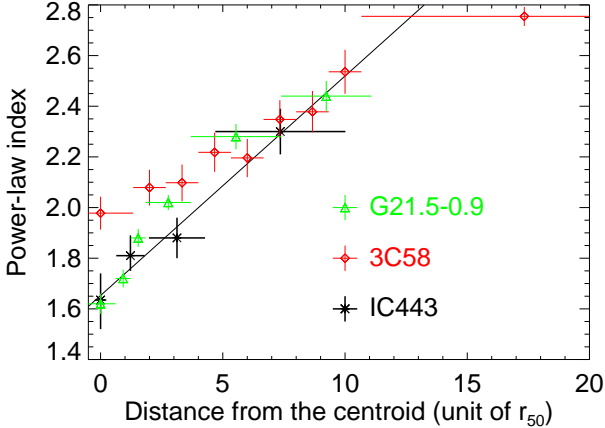,width=9.0cm}}
  \caption{$\gamma-r$ relation in the IC~443 plerion nebula (this
  paper), 3C~58 (\protect\cite{bwm01}) and G21.5-0.9
  (\protect\cite{wbb01}). The X-axis shows the weighted mean
  distance of the pixels of a given region from the centroid of the
  nebula, expressed in unit of $r_{50}$, the radius at which the
  plerion surface brightness drop by a factor of 2 (12\arcsec,
  6\arcsec\  and 13\arcsec\  for IC~443, 3C~58 and G21.5-0.9,
  respectively).  Using the $r_{50}$ unit gives a measure
independent on the distance to the nebulae, while $r_{50}$ can be
easily converted into angular or absolute distances using the SNR
distances and above angular measures of $r_{50}$.  The
  positional error bar are the standard deviation of the distances. A
  fit with a linear relation for IC~443 only is also plotted}

\label{rgamma} 
\end{figure}

The results of the spectral analysis are shown in Table~\ref{srcpar} and
spectrum of the central core is shown in Fig.~\ref{plsp}. All the fits
provide a good description of the data. In Fig.~\ref{rgamma} we
present the spectral softening in the nebula going from the inner to
the outer regions, as well as the softening observed in other plerions,
for comparison.
Letting the absorption value vary does not yield a better
fit, therefore we argue that there is no evidence for
any absorption variation across the plerion nebula.

%The observed softening can be fitted with a linear
%relation, $\gamma=(1.65\pm0.03)+(7.2\pm0.7) \times 10^{-3}r$, where $r$
%is in arcsec. 

\section{Discussion}

The \xmm\ observations allow us to resolve the structure of the hard X-ray
source \src. The observed power-law index steepening (Fig.~\ref{rgamma}) 
is consistent with the synchrotron burn-off effect
(\cite{kc84}) and remarkably similar to the effect observed in other
plerions, thus confirming the pulsar wind nebula interpretation
suggested by \cite*{ocw01}.  Note that the frequency index steepening
is consistent with the standard value of 0.5 from the one-zone model by
\cite*{che00}, and that the hard spectral index of the core suggests a
low X-ray efficiency, similar to the Vela-X nebula and CTB~80 as
reported by \cite*{che00}.  Following the one-zone Chevalier (2000),
away from cooling regime the $L_{\rm X}/\dot{E}$ ratio is dependent on the
fraction $\epsilon_{\rm B}$ of energy density of the emitting region going into
magnetic fields ($L_{\rm X}/\dot{E} \propto \epsilon_{\rm B}^{0.83}$), and this
quantity is rather uncertain.  However, assuming $L_{\rm X}/\dot{E}=0.002$,
like in the bow-shock nebula CTB~80, and using the sum of the flux in
all the elliptical regions ($2.6\times 10^{33}$ erg s$^{-1}$ in the
0.2--4 keV band at 1.5 kpc), we obtain $\dot{E}=1.3\times 10^{36}$
\ergsec\  ($0.8\times 10^{36}$ \ergsec\  with the empirical relation of
\cite{sw88}), similar to the value adopted by \cite*{ocw01}.

Another important characteristic of this plerion nebula is that its
X-ray extent appears larger then its radio extension, an effect also
observed in G21.5-0.9 (\cite{wbb01}) and 3C~58 (\cite{bwm01}).  This is
unexpected if both X-ray and radio are just the synchrotron emission
of the nebula. However, the inverse Compton and non-thermal
bremsstrahlung components of the X-ray emission, as well as other thermal
components related to a possible shell, may become important in the
outer halo of the nebula.  Unfortunately, there is no way to verify if,
apart from the contamination of the IC~443 thermal emission, the
spectrum of the outer nebula is purely non-thermal (like in G21.5-0.9,
\cite{wbb01}), or contains an additional thermal component due to the
expansion of the nebula into the ejecta (as in 3C~58, \cite{bwm01}). On
the other hand, such a small radio nebula may be the result of a low
sensitivity of the VLA observations reported by \cite*{ocw01}. The
X-ray to radio luminosity ratio, defined for instance as in
\cite*{wsp97}, is only a factor of 2 less then the Crab's and 100 more
than 3C~58. But if deeper radio observations reveal a nebula
comparable with, or larger then, the X-ray nebula the real value of the ratio
may be reduced.

By extrapolating back the sum of the spectra of the different nebula
regions in the radio regime, and comparing them with the radio spectrum
reported by \cite*{ocw01}, we have found a spectral break at 100
GHz, instead of 10,000 GHz as reported by \cite*{ocw01}. The
difference is due to the contribution of the hard inner nebula regions, which
were not properly modeled using ASCA data, on which the \cite*{ocw01}
estimation relies upom. Such a low frequency break is also found in
3C~58 and G21.5-0.9, as well as in other examples of other ``non
Crab-like" plerion reported by \cite*{wsp97}.

It should be noted, that there are some differences between the IC~443
plerion and 3C~58 (and G21.5-0.9). In particular, the spectral index
steepening in 3C~58 and G21.5-0.9 is higher than 0.5, the value
predicted by the one-zone model of \cite*{che00}.  Another key
difference is the central source in the center, which is very prominent
in case of the IC~443 plerion. However, on the basis of the published
Chandra data on G21.5-0.9 (\cite{scs00}), the presence of central
compact sources in G21.5-0.9 cannot be excluded.

As for the NS, a better interpretation of the its temperature
constraint requires a careful analysis of the NS atmosphere effect
(e.g., \cite{psz95}) and general relativitistic corrections.  The effect of
the hydrogen atmosphere could reduce the effective temperature
typically by a factor 1.3 to 2 and thus could make the effective NS
radius consistent with the standard NS radius of $\sim$10 km. It is worth
noting that the NS temperature of $kT \sim$ 0.1 keV is nicely
consistent with the IC~443 age of 30,000 years (\cite{che99}), if the
standard NS cooling curve of superfluid NS is used (e.g., \cite{ykl99},
- see their Fig.~7), but a lower age cannot be excluded.  The
discrepancy between our best guess for the NS surface temperature and
the temperature derived by \cite*{ocw01} by fitting the point source
CXOU~J061705.3+222127 may be due to the low value of $N_{\rm H}$ assumed by
them ($1.3\times 10^{21}$ cm$^{-2}$). We verified that this $N_{\rm H}$ value
is rejected at 99\% confidence level by all our fits. A deeper
\ch\ observation will better constrain the physical properties of the
compact source.

If the hard power-law spectrum of the nebula core region is
extrapolated up to GeV regime, it would provide a flux of 2.0
(0.4--15.2)$\times 10^{-7}$ ph cm$^{-2}$ s$^{-1}$, which is consistent
with the {\it EGRET} flux of 3EG~J0617+2238 ($5.0\pm 0.4\times 10^{-7}$
ph cm$^{-2}$ s$^{-1}$).  The nebula is placed outside the 95\%  uncertainty
circle of the 3EG~J0617+2238, but the systematic uncertainties of the
{\it EGRET} circle allows us to speculate about possible association of
the two sources. If, however, the GeV regime {\it
EGRET} photons indeed originate from synchrotron emission of
relativistic electrons, this would seriously constrain the nebula model
of \cite*{kc84} where e$^{\pm}$ are accelerated by a relativistic wind
termination shock and then radiate in the downstream region.  The
maximum photon energy $\varepsilon$ (measured in GeV) produced by an
accelerated electron of maximal Lorentz factor ${\rm \Gamma_{\rm max}}$ in the
magnetic field of the emitting region B${\rm _e}$ (measured in G) can be
obtained from $8.6 \times 10^{16} \varepsilon \approx B_{\rm e}
\Gamma_{\rm max}^2$ G.  The e$^{\pm}$ synchrotron loss time, ${\rm t_s \approx
5\cdot 10^8 B_{\rm a}^{-2} \Gamma^{-1}}$ s, must exceed  the e$^{\pm}$
acceleration time by relativistic shock, t${\rm _a}$ (where ${\rm B_a}$ is the
magnetic field in the acceleration region).  Since ${\rm t_a}$ cannot be less
then the particle gyroperiod, one may obtain ${\rm B_a \Gamma_{max}^2 \ll
10^{16}}$~G. Thus, the magnetic field amplification in the shock
downstream region  ${\rm B_e \gg 8 B_a \varepsilon}$ is required.  The
amplification effect can be studied with high spatial resolution X-ray
imaging. The EGRET luminosity of $\sim 10^{35}$ \ergsec\  would not be a
negligible fraction of the spin-down power estimated from the
\cite*{sw88} relation. An alternative explanation for the {\it EGRET}
emission is the relativistic bremsstrahlung of radio emitting electrons
accelerated by the supernova radiative shock (\cite{bce00}), and future
INTEGRAL and GLAST observations will resolve that alternative, as well
as providing an accurate position for 3EG~J0617+2238.

\section{Summary and conclusions}

We report \xmm\  Cal/PV observations of the hard X-ray nebula inside
the supernova remnant IC~443, recently discovered by \ch\
(\cite{ocw01}).  The longer exposure time and the high \xmm\  effective
area have allowed us to trace and to study the diffuse emission of the
nebula far beyond the extension reported by \cite*{ocw01}. We confirm
the plerionic nature of the nebula, as suggested by \cite*{ocw01}, by
virtue of the observation of the synchrotron burn-off, which is
in agreement with plerion model expectations. The
relatively hard power-law index of the core, the flat radio spectrum,
and the X-ray extension much larger then the radio extension at 8.46
GHz make this plerion more similar to 3C~58 and G21.5-0.9 (two well
known examples of ``non Crab-like" or ``second kind" plerions,
\cite{wsp97}), rather than to the Crab nebula. Our data allows us to
place only upper-limit on the presence of a thermal compact X-ray
source in the nebula core. However, we argue that its temperature
is likely be $\sim 0.1$ keV, significantly lower that the
temperature suggested by \cite*{ocw01}. We also argue that, if the
nebula is the counterpart of the {\it EGRET} source 3EG~J0617+2238, as not
excluded by the extrapolation of its X-ray spectrum, a strong constraint
on magnetic field amplification in the nebula may be posed.

\begin{acknowledgements}

F. Bocchino acknowledges an ESA Research Fellowship.  The work of A.M.B
was supported by the INTAS-ESA 99-1627 grant. We thank A. N. Parmar and
the anonymous referee for their useful comments on the manuscript. We
also thank the ESTEC \xmm\ staff for their suggestions about data
analysis.

\end{acknowledgements}

\bibliography{references}

\bibliographystyle{aabib}

\end{document}